\documentclass[11pt,a4paper]{article}

\usepackage[latin1]{inputenc}
\usepackage[english]{babel}
\usepackage{amsmath,amssymb}
\usepackage{graphicx}
\usepackage{cite}
\usepackage{place ins}

\newcommand{\as}{\alpha_s}

\newcommand{\beq}{\begin{equation}}
\newcommand{\eeq}{\end{equation}}
\newcommand{\bea}{\begin{eqnarray}}
\newcommand{\eea}{\end{eqnarray}}
\newcommand{\bdm}{\begin{displaymath}}
\newcommand{\edm}{\end{displaymath}}
\def\as{\alpha_s}

\def\ord{{\cal O}}

\def\d{\partial}

\def \d{{\rm d} }
\def \d0 {D\O \;}

\begin{document}
\begin{titlepage}
\begin{flushright}
{DCPT/12/66 \\ IPPP/12/33\\MAN/HEP/2012/02}
\end{flushright}     

\begin{center}
{\Large \bf
Predictions for Drell-Yan $\phi^*$ and $Q_T$ observables at the LHC}
\vspace*{1.5cm}

Andrea Banfi$^{a}$, Mrinal Dasgupta$^b$, Simone Marzani$^c$ and Lee~Tomlinson$^d$
\\
\vspace{0.3cm}  {\it
{}$^a$Physikalisches Institut, Albert-Ludwigs-Universit\"at Freiburg,\\ D-79104 Freiburg, Germany \\
{}$^b$Consortium for Fundamental Physics, \\School of Physics \& Astronomy, University of Manchester, \\ Manchester M13 9PL, United Kingdom\\
 {}$^c$Institute for Particle Physics Phenomenology, Durham University,\\ Durham DH1 3LE, United Kingdom\\
{}$^d$School of Physics \& Astronomy, University of Manchester, \\ Manchester M13 9PL, United Kingdom}
\vspace*{1.5cm}

\bigskip
\bigskip

 { \bf Abstract }
\end{center}
\begin{quote}

We make theoretical predictions for the recently introduced variable $\phi^*$ corresponding to the azimuthal angle between leptons produced in the Drell-Yan process at the LHC. As a consequence of this work we are also able to generate 
results for the more commonly studied transverse momentum $Q_T$ of the lepton 
pair. Comparisons of these purely perturbative estimates for the $Q_T$ case 
yield good agreement with ATLAS and CMS data, as we demonstrate. We anticipate that this work will help stimulate measurements of $\phi^*$ at the LHC.

\end{quote}
\end{titlepage}

\section{Introduction}
\label{sec:introduction}

The transverse momentum ($Q_T$) distribution of lepton pairs produced in hadron-hadron collisions is one of the most studied observables in particle phenomenology. Perturbative QCD corrections are under control at next-to-leading order (NLO) and fully differential codes to generate them are publicly available, e.g.~\cite{mcfm,FEWZ, FEWZ2,FEWZ2_1,dynnlo}. However, fixed-order predictions fail to describe the small-$Q_T$ region and resummation techniques must be used in order to obtain reliable predictions. The resummation of large logarithms in the $Q_T$ spectrum has been studied by several groups~\cite{DDT, DSW,APP,CSS,ERV,EV,CdFGuniv,GrazzDeFlo,GrazziniDeFlorian,FlorenceQT,Bozzipt, FlorenceDY,BecherNeubertA3,BecherNeubertPheno}. The state of the art in the resummation is next-to--next-to-leading logarithmic accuracy (NNLL) in the resummed exponent~\cite{FlorenceQT}.

On the experimental side, the transverse momentum distribution of $Z$ bosons decaying to lepton pairs has been intensively studied by the \d0 and CDF collaborations at the Tevatron~\cite{CDFRunI, D0RunI, D0RunII}. More recently, measurements have been performed at the LHC as well~\cite{atlasZpt,cmsZpt}.

State-of-the-art theoretical calculations have been compared to data coming 
from the Tevatron experiments with the aim of testing QCD predictions and to assess the importance of non-perturbative physics. For instance, in Ref.~\cite{FlorenceDY} a NNLL resummation, matched to NLO calculation was found to provide a good description of Tevatron Run II data, within theoretical and experimental uncertainties, with no apparent need for non-perturbative contributions.  A somewhat different approach is used in the program RESBOS~\cite{resbos, resbos_comp}. In this case the evaluation of the theoretical uncertainty that affects the resummation is less thorough and the agreement with the data is restored thanks to a non-perturbative contribution~\cite{resbos_comp, D0RunII}. In our view, the difference between the pure perturbative result for the $Q_T$ spectrum and the one with non-perturbative corrections included can be similar in size to the perturbative uncertainty. We would thus prefer to evaluate the perturbative uncertainty carefully {\emph{prior}} to including non-perturbative effects via a 
phenomenological study in conjunction with experimental data.

Further obscuring the picture is the experimental resolution which affects the measurement of transverse momenta. For this reason novel variables have been introduced in~\cite{WV,WVBRW} and measured by the D\O $ $ collaboration~\cite{D0dphi}. These variables, labelled $a_T$ and $\phi^*$, both crucially depend on 
the azimuthal angle $\Delta \phi$ between the final state leptons, at low $Q_T$. The experimental resolution for $a_T$ and $\phi^*$ is significantly better 
than the one for $Q_T$~\cite{WVBRW}, which enables the possibility to better test theoretical ideas and constrain non-perturbative effects. An example 
of such a phenomenological study with the RESBOS code can be found in Ref.~\cite{D0dphi}, where the data clearly disfavoured the inclusion of non-perturbative small-$x$ broadening terms included in the RESBOS code.

In previous papers \cite{BDDaT,BDMphi, BDMT} we  provided the details of a resummed treatment of these new variables and discussed their relationship to $Q_T$ and to each other. We computed the resummation to NNLL accuracy, carried out the matching to fixed-order NLO results from MCFM~\cite{mcfm} and performed a complete phenomenological study, including a faithful estimate of the theoretical uncertainty. We compared our prediction to the \d0 data and we found that resummed perturbation theory provides an excellent description of the $\phi^*$ distribution with little variation in the quality of our description in the different rapidity regions. 

In this Letter we extend our work by providing resummed and matched predictions for the LHC. We begin by considering the standard $Q_T$ spectrum and we compare our results to the measurements performed by ATLAS and CMS. Having thus obtained a validation of our method, we then provide theoretical predictions for the $\phi^*$ variable, which should stimulate its measurement at the LHC.

\section{Resummation of the $Q_T$ and $\phi^*$ distributions}
The resummation formalism for the Drell-Yan $Q_T$ spectrum is well established. The resummation can be performed to NNLL accuracy~\cite{FlorenceQT} and it can be matched to NLO calculations to obtain accurate predictions over a vast range of $Q_T$.

Here we merely  remind the reader that in order to properly treat transverse momentum conservation in the case of $n$-gluon emissions with transverse momentum $\underline{k}_{Ti}$, one usually introduces a two-dimensional impact parameter $\underline{b}$, which is the Fourier conjugate of the lepton-pair transverse momentum $\underline{Q}_T$. The delta function which enforces momentum conservation in the transverse plane can be then expressed in a factorised form:
\beq
\delta^{(2)}\left(\sum_{i=1}^n \underline{k}_{Ti}+\underline{Q}_T \right)=\frac{1}{4 \pi^2}\int {\rm d} ^2 \underline{b} \, e^{i\underline{b}\cdot \underline{Q}_T}\prod_{i=1}^n e^{i\underline{b}\cdot \underline{k}_{Ti}}.
\eeq
If one is interested in the $Q_T$ spectrum one integrates over the angle between $\underline{b}$ and $\underline{Q}_T$, obtaining a Bessel function. The resummed differential distribution has the form
\bea \label{eq:resummedQT}
\frac{ {\rm d} \sigma}{ {\rm d} Q_T} \left( Q_T,M, \cos \theta^*, y \right) &=& \frac{\pi \alpha^2}{s N_c} 
\int_0^{\infty} {\rm d}  b\, b \,   Q_T\, J_0\left(b Q_T \right) 
e^{-R(\bar{b},M, \mu_Q,\mu_R)} 
 \nonumber \\ &&\times\, \Sigma \left(x_1,x_2,\cos\theta^*, b,M,\mu_Q,\mu_R,\mu_F \right)\,,
\eea
where
$x_{1,2} = \frac{M}{\sqrt{s}}e^{\pm y}$, being $M$ and $y$ the dilepton's invariant mass and rapidity, respectively, and $\bar{b}= \frac{b e^{\gamma_E}}{2}$.
The resummed result depends on three arbitrary scales: renormalisation and factorisation scales, $\mu_R$ and $\mu_F$, as well as the resummation scale $\mu_Q$. Variations of these scales around the dilepton invariant mass $M$ provide us with an estimate of the theoretical uncertainty.
In Eq.~(\ref{eq:resummedQT}) the dependence upon the large logarithms we wish to resum is encoded in the radiator:
\bea\label{radiator}
R\left(\bar{b}\mu_Q,\frac{M}{\mu_Q},\frac{\mu_Q}{\mu_R};\as(\mu_R)\right) &=& L g^{(1)}(\as L) + g^{(2)}\left(\as L, \frac{M}{\mu_Q},\frac{\mu_Q}{\mu_R}\right) + \frac{\as}{\pi} g^{(3)}\left(\as L, \frac{M}{\mu_Q},\frac{\mu_Q}{\mu_R}\right)\,, \nonumber \\ 
\eea
where $L=\ln(\bar{b}^2\mu_Q^2)$ and $\as = \as(\mu_R)$,  as well as in the cross-section $\Sigma$, via DGLAP evolution of the parton distribution functions from the hard scale of the process to $Q=\frac{\mu_F}{\mu_Q \bar{b}}$.
We refer the reader to the Appendices of our previous work~\cite{BDMT} for the explicit expressions of the functions $g^{(i)}$ as well as for the cross-section $\Sigma$~\footnote{Our expression for $\Sigma$ does not include the $\ord\left(\as^2 \right)$ correction which is in principle needed to achieve complete NNLL accuracy.} . It is well known that the $b$-integral in Eq.~(\ref{eq:resummedQT}) has issues both at large and small $b$. We evaluate the integral by introducing an upper cut-off $b_{\rm max}$ and by freezing the radiator for  $b <b_{\rm min}$, as explained in detail in Ref.~\cite{BDMT}. 

An issue that has become apparent in the current study is related to the behaviour of the parton distribution functions at low momentum or, equivalently, at large $b$. 
In our Tevatron study~\cite{BDMT} we decided to freeze the parton densities below $Q_0=1$~GeV. We tested this prescription by varying the value of $Q_0$ within a factor of two, finding that the sensitivity was much less than the perturbative uncertainty.  However, at the LHC energies, the parton distribution functions $f_i(x,Q)$ are typically probed at lower values of $x$, where the $Q$ dependence is steeper. In the resummed calculation the argument of the parton densities is $Q=\frac{\mu_F}{\mu_Q \bar{b}}$; we have noticed that for low values of the ratio $\mu_F/\mu_Q$ the introduction of an abrupt freezing point at $1$~GeV gives rise to an oscillatory behaviour of the $Q_T$-distribution. 
We circumvent this issue by constructing an exponential extrapolation of the parton densities below $1$~GeV. This procedure cures the oscillations but still produces curves with a peak which is noticeably shifted to the right with respect to our central value $\mu_Q=\mu_F=\mu_R=M$. Such curves that critically depend on the behaviour of the parton densities below $1$~GeV are not reliable in our framework based on collinear factorisation and DGLAP evolution of collinear parton distribution functions. It is far from obvious that one can trust this framework at very low scales, where small-$x$ effects should be taken into account and collinear factorisation might not be valid, signalling the need for transverse momentum dependent parton distributions~\cite{cch, CollinsEllis, ACQR}. 
It would be interesting to see whether the sensitivity we observe, when we vary all scales independently, is also present in different theoretical implementations. However, only the approach of Refs~\cite{CdFGuniv,GrazzDeFlo,GrazziniDeFlorian,FlorenceQT,Bozzipt, FlorenceDY} takes into account independent scale variation, as we do, and, to our knowledge, results at LHC energies have been published by that group for the Higgs $Q_T$ distribution but not for the $Z$ transverse momentum.

In this study we adopt a more phenomenological viewpoint. We note that if we vary the three scales ($\mu_Q,\mu_R$ and $\mu_F$) independently, all the curves for which $\frac{\mu_F}{\mu_Q}\ge1$ group together around the central one, while the ones for which $\frac{\mu_F}{\mu_Q}=\frac{1}{2}$ tend to form a distinctly outlying family of curves, with a shifted peak. Therefore, we decide to vary the perturbative scales independently, with the additional constraint $\frac{\mu_F}{\mu_Q}\ge1$, taking the resulting band as an estimate of the perturbative uncertainty of our resummed prediction. Hence the curves we produce are indeed insensitive to the behaviour of the parton distributions at very low scales. 

In this Letter we are also considering the variable $\phi^*$, which is defined as~\cite{WVBRW}
\begin{equation}\label{phistardef}
\phi^* = \tan \left (\frac{\pi-\Delta \phi}{2} \right) \sin \theta^* = \left | \sum_i \frac{k_{Ti}}{M} \sin \phi_i \right | +\ord\left(\frac{k_{Ti}^2}{M^2}\right)\,,
\end{equation}
where $\Delta \phi$ is the azimuthal angle between the two leptons produced by the $Z/\gamma^*$ decay and $\sin \theta^*$ is the scattering angle of the dileptons with respect to the beam, in the boosted frame where the leptons are aligned. This definition of $\theta^*$ avoids the necessity of measuring magnitudes of lepton momenta and significantly helps the experimental resolution as explained in Ref.~\cite{WVBRW}.  We have also introduced the angles $\phi_i$ between the momentum of gluon $i$ with respect to the lepton axis in the transverse plane. 
Thus, $\phi^*$ is essentially determined by one component of the  transverse vector $\underline{Q}_T$. For this reason, the formalism of $Q_T$ resummation can be applied to the variable $\phi^*$ as well~\cite{BDDaT,BDMphi, BDMT}.
We can straightforwardly perform the integral over one component of $\underline{b}$, obtaining a cosine instead of the Bessel function
\bea
\label{eq:resummedphistar}
\frac{ {\rm d} \sigma}{ {\rm d} \phi^*} \left( \phi^*,M, \cos \theta^*, y \right) &=& \frac{\pi \alpha^2}{s N_c} 
\int_0^{\infty} {\rm d} b\, M \,\cos \left(bM \phi^* \right) 
e^{-R(\bar{b},M, \mu_Q,\mu_R)} \nonumber \\ &&\times\,\Sigma \left(x_1,x_2,\cos\theta^*, b,M,\mu_Q,\mu_R,\mu_F \right)\,.
\eea

 The theoretical prediction for the $Q_T$ or $\phi^*$ distributions is obtained by matching the resummed calculation to a fixed order computed with the program MCFM~\cite{mcfm}:
\begin{equation} \label{matched}
\left(\frac{ {\rm d} \sigma}{{\rm d} v}\right)_{\mathrm{matched}} = \left(\frac{{\rm d} \sigma}{ {\rm d} v}\right)_{\mathrm{NNLL }} +\left(\frac{{\rm d} \sigma}{{\rm d} v}\right)_{\mathrm{NLO}}-\left(\frac{{\rm d} \sigma}{{\rm d} v}\right)_{\mathrm{expanded}}\,,
\end{equation}
with $v=Q_T, \phi^*$.
The last term in the above equation is the expansion of the resummation to $\ord \left( \as^2 \right)$ and avoids double counting. 
We note that the difference between the resummation and its expansion is well behaved only up to values of $Q_T$ of the order of the $Z$ mass. This can be linked to the fact that resummation formalisms in impact parameter space suffer from a non-physical behaviour at sufficiently large transverse momenta~\cite{FlorenceDY}. Therefore, for $Q_T \gtrsim m_Z$ our result is the pure NLO.
The expressions in Eqs.~(\ref{eq:resummedQT}),~(\ref{eq:resummedphistar}), as well as the fixed-order prediction from MCFM, are fully differential in the dilepton kinematics, so that we can take into account any experimental cuts.
 \FloatBarrier
 
\section{The $Q_T$ and $\phi^*$ distributions at the LHC}
\begin{figure}
\begin{center}
\includegraphics[width=0.495 \textwidth]{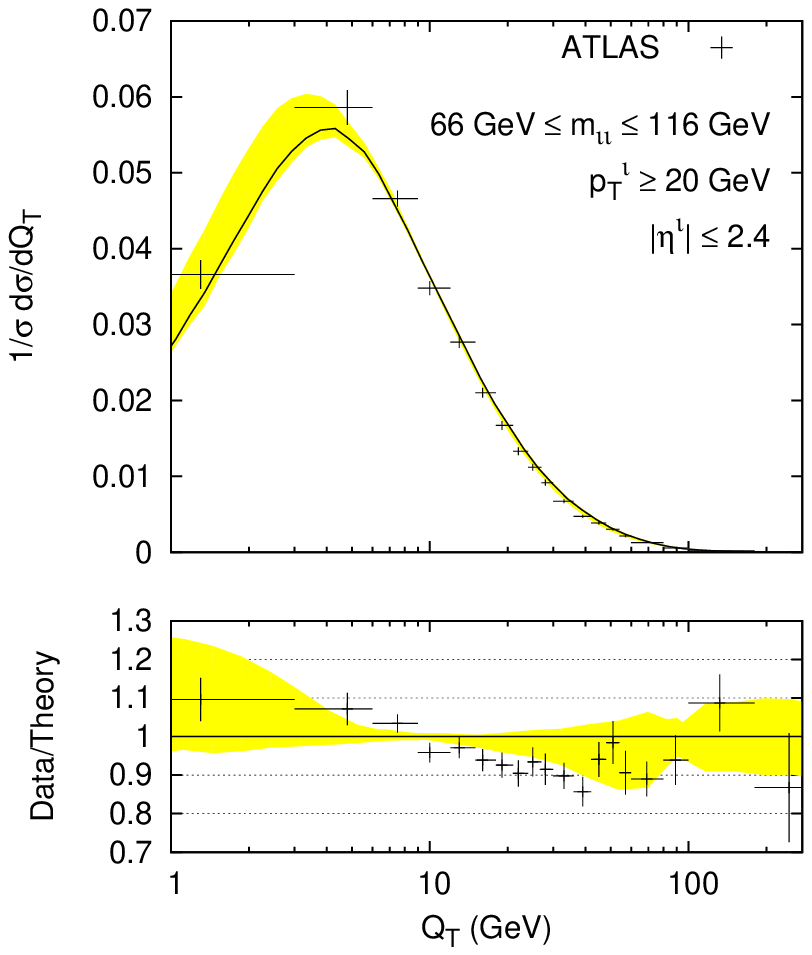}
\includegraphics[width=0.495 \textwidth]{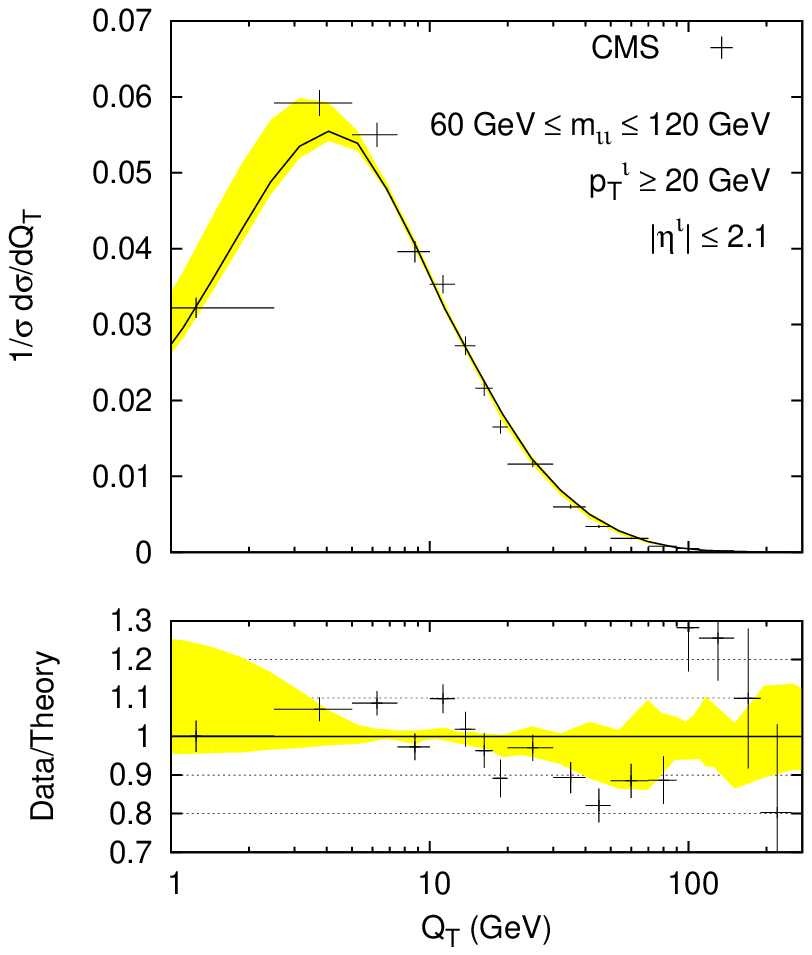}

\caption{Comparison of the theoretical prediction NNLL+NLO for the $Q_T$ spectrum to the experimental data collected by the ATLAS collaboration (on the left) and by the CMS collaboration (on the right). The theoretical uncertainty is obtained by varying the perturbative scales, as explained in the text. The curves are normalised to unit area, as the data are.}
\label{fig:QTlhc}
\end{center}
\end{figure}
In this section we compare our resummed and matched results to the measurement of the $Z/\gamma^*$ transverse momentum distribution in  proton-proton collisions at $7$~TeV, performed by the ATLAS and CMS collaborations. The cuts applied by the two experiments slightly differ. 
We stress once more that because our predictions are fully differential in the lepton momenta we can take these cuts into account, without any need to extrapolate to the full lepton acceptance. Performing comparisons to the data in the fiducial region is very important if one wants to fully exploit the potential of the measurement, as recently pointed out, for instance, in Ref~\cite{FEWZ2_1} in the context of $W^{\pm}$ production and studies of parton distribution functions.

In particular, we compare our theoretical prediction to the ATLAS data (electrons and muons) in the fiducial volume, defined by $p_T^l>20$~GeV and $|\eta^l|<2.4$, in the lepton-pair invariant mass window $66<m_{ll}<116$~GeV.  The CMS data instead are presented for  $p_T^l>20$~GeV, $|\eta^l|<2.1$ and $60<m_{ll}<120$~GeV.
In Figure~\ref{fig:QTlhc} we compare our theoretical prediction to the ATLAS and CMS data. The theoretical curve is computed using the CTEQ6m set of parton densities~\cite{cteq6m}, with the value of the strong coupling taken from the fit, $\as(M_Z)=0.1179$. The curve is normalised  to its own area. The band represents our assessment of the perturbative uncertainty. It is obtained by varying the perturbative scales $\mu_Q$, $\mu_F$ and $\mu_R$ between $M/2$ and $2 M$, with the constraints $\frac{1}{2}\le\frac{\mu_i}{\mu_j}\le2$ and $\frac{\mu_F}{\mu_Q}\ge1$,  where $i,j \in \{F, Q, R\}$. From this we can estimate the perturbative uncertainty to be around $10$\% in the peak region we are most interested in.
Another source of theoretical uncertainty comes from the parton distribution functions. We estimate this one to be smaller (of the order 2\% at low $Q_T$) than the one obtained with scale variation. This is expected because we are considering the shape of the $Q_T$ distribution and the uncertainty coming from the parton densities largely cancels when we perform the ratio to the inclusive cross-section. Our predictions are computed without any explicit non-perturbative effects. We have checked that the inclusion of a Gaussian contribution in $b$-space with coefficient $g_{\rm NP}=0.5$~GeV$^{2}$, taken from our Tevatron study~\cite{BDMT},  produces distributions compatible with our perturbative uncertainty.

The ratio plots show that we have a good description of the data in the low-$Q_T$ region, within theoretical and experimental uncertainty. For intermediate values of the transverse momentum our prediction slightly overshoots the data points from both experiments, by 5\% to 10\%. In the large $Q_T$ region, where our prediction reduces to NLO, we correctly describe the ATLAS data. The agreement with the CMS data is instead not as good in this region; this feature is also present in the comparison to the NLO distribution performed by the collaboration itself~\cite{cmsZpt}.

The experimental collaborations compared their data~\cite{atlasZpt,cmsZpt} to the fixed-order prediction computed with FEWZ~\cite{FEWZ} and to various Monte Carlo parton showers with or without interfaces with NLO codes, e.g. POWHEG~\cite{powheg}. As expected, fixed-order perturbation theory works reasonably well at high-$Q_T$ but resummation is needed in the low-$Q_T$ region. Parton showers and RESBOS~\cite{resbos} were found to provide very good descriptions of the data, although the main source of theoretical uncertainty at low-$Q_T$, the resummation scale $\mu_Q$, was not explored in those studies. Furthermore, both parton showers and RESBOS predictions contain non-perturbative effects.

Finally, a comparison between the theory prediction using SCET and the ATLAS data have been performed in~\cite{BecherNeubertPheno} in the intermediate-to-low-$Q_T$ region. Good agreement was found in the study with the introduction of a non-perturbative contribution to describe the lowest-$Q_T$ data points.

\begin{figure}
\begin{center}
\includegraphics[width=0.49 \textwidth]{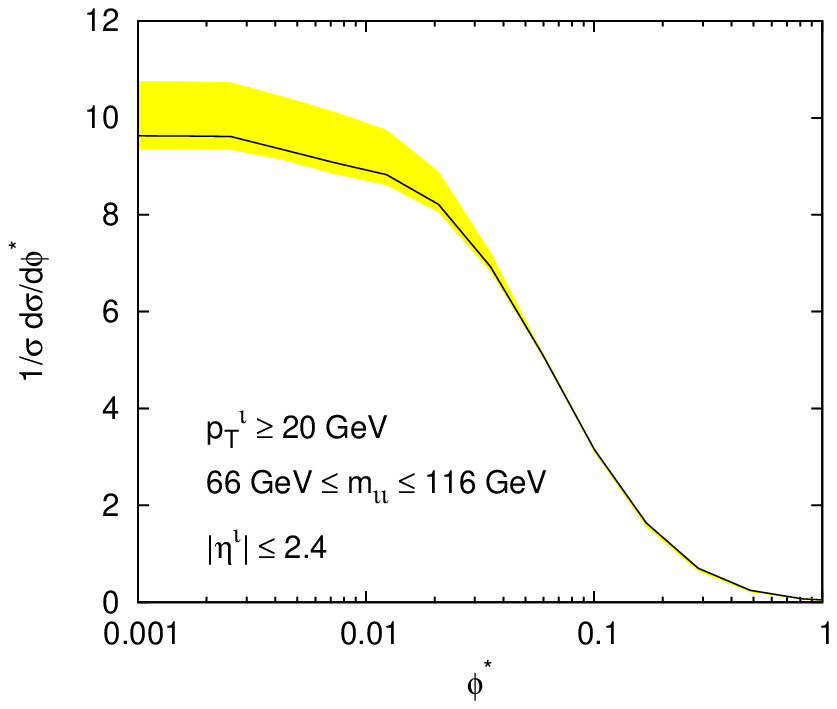}
\includegraphics[width=0.49 \textwidth]{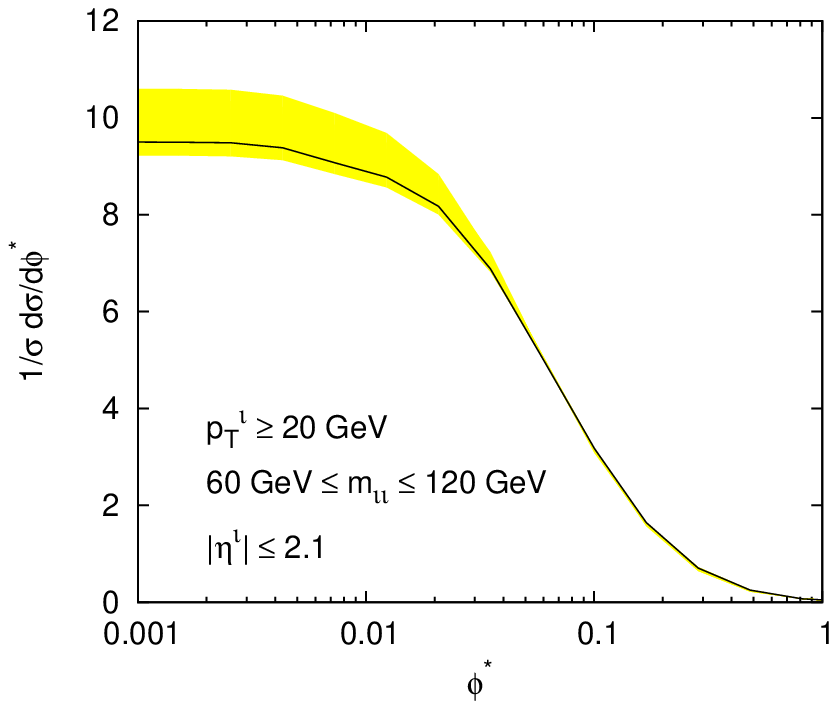}
\caption{Theoretical predictions at  NNLL+NLO for the normalised $\phi^*$ distribution in $pp$ collisions at $7$~TeV. The kinematical cuts are the same as the ones adopted for the $Q_T$ case by the ATLAS collaboration (on the left) and by the CMS collaboration (on the right).}
\label{fig:phistar}
\end{center}
\end{figure}

Having validated our resummation procedure at the LHC energies, we can now  provide a prediction for the $\phi^*$ distribution in proton-proton collision at $7$~TeV. We nominally choose the same selection cuts as the ones adopted by the ATLAS and CMS collaborations for $Q_T$. The resummed and matched results are shown in Fig.~\ref{fig:phistar}, in the case of ATLAS cuts (on the left) and CMS cuts (on the right). The theoretical uncertainty is estimated as explained above, and in the low-$\phi^*$ region it is $\ord(10\%)$. 
The size of the uncertainty band is comparable to the one we found for this distribution at the Tevatron. Furthermore, the height of the plateau is significantly reduced. This is due to fact that more radiation is produced when the parton distribution functions are probed at lower values of $x$, with the result that  configurations with Born kinematics are less likely.

We advocate the measurement of the $\phi^*$ distribution at the LHC, in order to accurately probe the low-$Q_T$ domain of lepton pairs produced via the Drell-Yan mechanism. Moreover, the LHC experiments can probe different regions of phase-space such as, for instance, forward rapidities or low values of the dilepton invariant mass. In these regions one expects small-$x$ contributions, which  go beyond the resummation formalism currently used,  to become important, signalling the need for better theoretical modelling~\cite{BM_dy,fowDY,LipatovZotov,HautJung}.

\section{Conclusions}
In this Letter we have compared a resummed and matched prediction for the $Q_T$ spectrum of $Z$ boson in proton-proton collisions at $7$~TeV. The accuracy of the resummation is NNLL and it is matched to a NLO calculation obtained with the program MCFM~\cite{mcfm}. Our prediction is fully differential in the leptons' momenta and therefore can be compared directly to the data in the fiducial region.  
We have found good agreement between the experimental data and our calculation in the low-$Q_T$ region we are most interested in even without direct inclusion of non-perturbative effects, which is a similar finding to that of our previous phenomenological study of $\phi^*$ at the Tevatron \cite{BDMT}.
Additionally in this Letter we have provided the theoretical prediction for the $\phi^*$ variable at the LHC.

Our resummed predictions are obtained with a computer program that we plan to release soon. The code can be easily interfaced with a fixed-order program (MCFM in this case), which provides the Born-level results. Our program then computes the desired observable in terms of the final states' kinematics and re-weights the Born cross-section with the resummation.
For this reason the numerical code can be easily applied to different processes, which involve a colour singlet in the final state. For instance, the resummation for separation in azimuth of the final state leptons $\Delta \phi$ can be immediately obtained. Moreover, because the resummation is blind to the actual nature of the (colour-singlet) final state, the same code can be applied to compute resummed prediction for processes with different final states, e.g. the azimuthal separation between vector bosons or Higgs and $Z$ bosons in associated Higgs production. The inclusion of gluon-induced processes, e.g. Higgs $Q_T$-spectrum or azimuthal diphoton distributions,  is also possible but it requires modifications of the resummed exponent. However, new structures arise if one wants to go beyond the leading-logarithmic accuracy~\cite{CatGrazSpin}. Finally, the methods we have used here are also applicable to the study of azimuthal angle between coloured particles in the final state (jets in dijet production) which would also be interesting to explore in detail phenomenologically~\cite{BanDasDel}.
\vspace{0.3 cm}

  {\bf Acknowledgments}
This work is supported by UK's STFC. The work of MD is supported by the Lancaster-Manchester-Sheffield Consortium for Fundamental Physics, under STFC grant ST/J000418/1. SM wishes to thank Richard Ball and Frank Krauss for useful discussions.

\end{document}